\documentclass[useAMS,usenatbib]{mn2e}

\usepackage[latin1]{inputenc}
\usepackage{graphicx}
\usepackage{epstopdf}

\usepackage{rotate}
\usepackage{amsmath, float}
\usepackage{amssymb}
\usepackage{soul}
\usepackage{txfonts}
\usepackage{fleqn}
\usepackage{color}
\usepackage{comment}

\usepackage{mathrsfs}  
\usepackage{booktabs}
\usepackage{colortbl} 

\newcolumntype{G}{>{\columncolor[gray]{0.8}}l} 

\newcommand{\be}{\begin{equation}}
\newcommand{\ee}{\end{equation}}
\newcommand{\bdm}{\begin{displaymath}}
\newcommand{\edm}{\end{displaymath}}
\newcommand{\bea}{\begin{multline}}
\newcommand{\eea}{\end{multline}}
\newcommand{\ba}{\begin{align}}
\newcommand{\ea}{\end{align}}

\def\simlt{\mathrel{\hbox{\rlap{\hbox{\lower4pt\hbox{$\sim$}}}\hbox{$<$}}}}
\def\simgt{\mathrel{\hbox{\rlap{\hbox{\lower4pt\hbox{$\sim$}}}\hbox{$>$}}}}

\title[Fe line in Scalar Tensor Theories]{Iron Line from Neutron Star
  Accretion
  Disks in Scalar Tensor Theories}
\author[N. Bucciantini \& J.Soldateschi]{
  Niccol\`o Bucciantini$^{1,2,3}$\thanks{E-mail: niccolo@arcetri.astro.it}
\&   Jacopo Soldateschi$^{2,1,3}$\thanks{E-mail: soldateschi@arcetri.astro.it}\\
$^{1}$INAF - Osservatorio Astrofisico di Arcetri, Largo E. Fermi 5,
I-50125 Firenze, Italy\\
$^{2}$Dipartimento di Fisica e Astronomia, Universit\`a degli Studi di Firenze, Via G. Sansone 1, 
I-50019 Sesto F.~no  (Firenze), Italy\\
$^{3}$INFN - Sezione di Firenze, Via G. Sansone 1, I-50019 Sesto F.~no  (Firenze), Italy}

\begin{document}
 
\date{Accepted / Received}

\maketitle

\label{firstpage}

\begin{abstract}
The Fe ${\rm K}_\alpha$ fluorescent line at $6.4$ keV is a powerful
probe of the space-time metric in the vicinity of accreting compact
objects. We investigated here how some alternative theories of
gravity, namely Scalar tensor Theories,
that invoke the presence of a non-minimally coupled scalar field and
predict the existence of strongly scalarized neutron stars, change the
expected line shape with respect to General Relativity. By taking into
account both deviations from the general relativistic orbital
dynamics of the accreting disk, where the Fe line originates, and the
changes in the light propagation around the neutron star, we computed
line shapes for various inclinations of the disk with respect to the
observer. We found that both the intensity of the low energy tails
and the position of the high energy edge of the line
change. Moreover we verified that even if those changes are in
general of the order of a few percent, they are potentially observable
with the next generation of X-ray satellites.

\end{abstract}


\section{Introduction}
\label{sec:intro}
Since the pioneering work of \citet{Brans_Dicke61a}, Scalar Tensor Theories
(STTs) have always attracted large attention from the scientific
community as viable gravitational extensions of General Relativity
(GR) \citep{Matsuda_Nariai73a,Damour_Esposito-Farese93a,Fujii_Maeda03a,Capozziello_de-Laurentis11a}, in part due to their mathematical simplicity, in
part because they seem to avoid many of the pathologies of other
extensions \citep{Papantonopoulos15a}. They key idea of STTs is to replace the
gravitational coupling constant $G$, with a dynamical non-minimally
coupled scalar field $\Phi$, such that the standard Einstein-Hilbert
Lagrangian is modified to:
\begin{align}
  \mathcal{S}_{\rm stt} =\frac{1}{16\pi}\int \sqrt{-\tilde{g}}[\Phi
  \tilde{R}-\frac{\omega(\Phi)}{\Phi}\tilde{\nabla}_\mu \Phi
  \tilde{\nabla}^\mu \Phi - 2\tilde{\Lambda}(\Phi)]d^4x,
  \label{eq:sttj}
\end{align}
where $\tilde{g}$ is the determinant of the metric tensor
$\tilde{g}_{\mu\nu}$, $\tilde{\nabla}_\mu$ the related covariant derivative,
$\tilde{R}$
the Ricci scalar, $\omega(\Phi)$ describes the field coupling and
$\tilde{\Lambda}(\Phi)$ is the scalar field potential.
The action of the scalar field $\Phi$ leads to non-linear behaviours
that in principle could account for cosmological observations, without the
need to invoke dark components \citep{Capozziello_de-Laurentis11a}.\\
\\
On top of the cosmological phenomenology associated to STTs, these
theories make interesting predictions also in the regime of strong
gravity, potentially leading to interesting deviations in the structure
and properties (e.g. the mass-radius relation) of neutron stars
(NSs) \citep{Damour_Esposito-Farese93a}. While solar-system experiments can set constraints on
the scalar field in the weak gravitational regime \citep{Shao_Sennett+17a}, only
compact objects can set limits in the strong one. Unfortunately, given that the no-hair theorem
forbids black holes (BHs) to
have any scalar charge \citep{Hawking72a}, the most promising environment to test gravity
in the strong regime, binary BH mergers, cannot be used to set any
constrain on the presence and nature of scalar fields \citep{Berti_Barausse+15a}. Only NSs offer
an environment compact enough for the scalar field to emerge. Indeed, some of these theories predict
that NSs can have sizeable scalar charges \citep{Damour_Esposito-Farese93a}, because of a
phenomenon known as \textit{spontaneous scalarization}. The presence of a scalar
field leads to new wave modes in binary NSs systems, beyond the standard
quadrupole gravitational wave emission. However, the present limits on
STTs based on the study of the orbital decay of binary pulsars
\citep{Shao_Sennett+17a,Anderson_Freire+19a} can be easily accommodated introducing screening
potentials or assuming massive scalar fields \citep{Yazadjiev_Doneva+16a}. On the other
hand it is not clear how, and how much a scalar field modifies the
final phases of binary NS inspiral before merger to a degree
observable with current instruments, and with specific signatures that
cannot be attributed to other causes (e.g. the equation of state). Even the measure of
the mass radius relation might not prove to be enough, if limited to few
objects, given its degeneracy with the equation of state. What we lack at
present is a way to probe deviations from GR in the close vicinity of
NSs. \\
One of the most powerful probes of the space time geometry close to
compact objects is light propagation. Light bending has been widely used in binary
pulsar systems \citep{Demorest_Pennucci+10a,Antoniadis_Freire+13a}, and more recently in the case of the BH at
the center of M87 \citep{Event-Horizon-Telescope-Collaboration_Akiyama+19a}. In accreting systems, one can also use
emission from the accretion disk, and in particular the shape of the
Fe ${\rm K}_\alpha$ fluorescent line at $6.4$ keV \citep{Miller07a,Dauser_Garcia+16a}. This line has been
extensively used in acceting BHs to measure their spin
\citep{Risaliti_Harrison+13a,Parker_Miller+18a,Kammoun_Nardini+18a}. Recently its has also been investigated in alternative
gravitational theories that predict deviation also for the BH
metric \citep{Yang_Ayzenberg+18a,Nampalliwar_Bambi+18a}. Despite the fact that this technique has been used just for
BHs, we know of many accreting NSs systems where
we observe the presence of this line \citep{Laor91a,Matt_Perola+92a,Degenaar_Miller+15a,Coughenour_Cackett+18a,Homan_Steiner+18a}. In principle Fe
${\rm K}_\alpha$ could be used to constrain the metric properties outside the
NS itself. \citet{Ghasemi-Nodehi18a} has shown how to
parametrize deviations from analytical solutions in GR, particularly relevant
for rapidly rotating NSs where the metric is only known numerically.  It has been suggested that the Fe line in accreting NSs
could be used to set limit on the NS radius, by modeling the effect on
the shape of the line due to the disk occultation by the surface of
the NS itself \citep{Cackett_Miller+08a,La-Placa_Stella+20a}. However in general these effects are found
to be small, of the order of few percents, and thus not measurable with
current instruments. They might in principle be within reach
of next generation X-ray satellites \citep{Barret_Lam-Trong+16a}. In the line
of \citet{Sotani17a}, who investigates light propagation from hot spots on the
surface of a scalarized NS, here we investigate how the Fe line
emission from an accreting disk around a NS is modified by the presence
of a scalar field with respect
to GR, including the effect of the possible occultation/truncation of the disk by
the NS itself.  In order to simplify the discussion, this paper is
mostly structured as a proof-of principle, and  not as a full fledged
investigation of the possible parameter space. For these reasons,
neither we compute realistic NS models based on physical equation of states,
nor we include rotation, and for the same reason we opted for the
simplest STT,  trying to parametrise the vacuum solution outside, in order to
provide a flexible estimate of the expected changes. 

This paper is organised as follows. In Sect.\ref{sec:stt} we briefly
review STTs, present the metric expected outside a scalarized
NSs, and introduce the ray-tracing. In  Sect.\ref{sec:ray} we
present and discuss the results both for unocculted and occulted
systems. We conclude in Sect.\ref{sec:conc}.

\section{Metric and ray-Tracing in Vacuum STT}
\label{sec:stt}

Depending on the nature and form of the coupling function $\omega(\Phi)$
or of the scalar field potential $\tilde{\Lambda}(\Phi)$  one can get different STTs, with
different phenomenologies. The simplest case is that of a mass-less
scalar field $\tilde{\Lambda}(\Phi)=0$. Varying the
action with respect to the metric and scalar field leads to a coupled
system of equations describing their mutual interplay. It can 
be easily  shown that such system, and in particular the generalisation of
Einstein field equations for the metric terms, contains higher order
derivatives, that change the mathematical nature of the equations
themselves \citep{Santiago_Silbergleit00a}.\\
It is possible however to recast the problem in terms of a
minimally-coupled scalar field $\mathcal{X}$, by performing a
conformal transformation from the original metric $\tilde{g}_{\mu\nu}$
to a new metric $\bar{g}_{\mu\nu} = \Phi
\tilde{g}_{\mu\nu}$. Expressed in terms of this new field and new
metric the Lagrangian reads:
\begin{align}
  \mathcal{S}_{\rm stt}  =\frac{1}{16\pi}\int \sqrt{-\bar{g}}[
  \bar{R}-2\bar{\nabla}_\mu \mathcal{X}
  \bar{\nabla}^\mu \mathcal{X} - 2\bar{\Lambda}(\mathcal{X})]d^4x,
    \label{eq:stte}
\end{align}
 where the bar indicates quantities relative to the new metric. This
 form leads to a set of field equations that are analogous to the standard
Einstein equations, supplemented with a well behaved momentum-energy tensor for
the scalar field. The original frame where the action read as in
Eq.~\ref{eq:sttj} is known as the  Jordan frame while the one where it
reads as  in Eq.~\ref{eq:stte} is known as the Einstein frame. The
relation between $\mathcal{X}$ and $\Phi$ is:
\begin{align}
\sqrt{2}\frac{{\rm d}\mathcal{X}}{{\rm d}\ln{\Phi}} = \sqrt{\omega(\Phi)+3/2}.
\label{eq:chitopsi}
\end {align}
In the simplest case of a mass-less scalar field,
the field equations in vacuum become:
\begin{align}
\bar{G}_{\mu\nu} = 2\bar{\nabla}_\mu \mathcal{X} \bar{\nabla}_\nu
  \mathcal{X} -\bar{g}_{\mu\nu}\bar{\nabla}_\kappa \mathcal{X} 
  \bar{\nabla}^\kappa \mathcal{X} ,
\end{align}
and
\begin{align}
 \bar{\nabla}_\mu  \bar{\nabla}^\mu \mathcal{X} = 0.
\end{align}
If one assumes steady state, $\partial_t =0$, and spherical symmetry (a
reasonable approximation for NSs not rotating close to the break-up
frequency), then it is possible to show that the line element in the Einstein
frame can be written in spherical coordinates $[r,\theta,\phi]$ as \citep{Just59a,Doneva_Yazadjiev+14a}:
\begin{align}
ds^2 = -f(r)^a dt^2 + f(r)^{-a} dr^2 + r^2 f(r)^{1-a} [d\theta^2
  +\sin^2{\theta} d\phi^2]
\end{align}
where the function $f(r)$ and the exponent $a$ depend on the total mass $M$ and scalar
charge $Q$ of the NS according to:
\begin{align}
f(r) = \left( 1- 2\sqrt{M^2+Q^2}/r \right),\quad a= M/\sqrt{M^2+Q^2},
\end{align}
while the scalar field is:
\begin{align}
\mathcal{X} = \frac{Q}{2\sqrt{M^2+Q^2}}\ln\left( 1- 2\sqrt{M^2+Q^2}/r \right).
\end{align}
However, in the Einstein frame, contrary to the Jordan frame, the weak
equivalence principle does not hold. In order to compute ray-tracing
using the standard geodesic equations, one needs to move back to the
Jordan frame and, to do so, to know the relation
between $\mathcal{X}$ and $\Phi$. One of the simplest possible choices
is to take  $\Phi = {\rm Exp}[-2\alpha_o\mathcal{X}
-\beta_o \mathcal{X}^2]$. $\alpha_o$ sets how strong deviations from GR
are in the weak field regime, and Solar system experiments constrain it
to be less than $\approx 10^{-4}$. $\beta_o$, on the other hand, sets how
strong scalarization effects can be in compact objects, and if smaller
than$\simeq -4$, it gives rise to strongly scalarized systems \citep{Will14a}. The Jordan metric
is then fully parametrised by the quantities $M$, $Q$, $\alpha_o$, and
$\beta_o$.\\
\\
If one makes the further assumption  $\alpha_o=0$, it is then possible
to derive an analytical expression for the Keplerian frequency of
matter orbiting the NS, using
the effective potential approach \citep{Abramowicz_Kluzniak05a,Doneva_Yazadjiev+14a}:
\begin{equation}
\Omega_{\rm k} = \frac{-\left( 1- \frac{2M}{ar} \right)^{2a}[\frac{2M^2}{a}+\beta_oQ^2
  \ln\left( 1-\frac{2M}{ar}\right)]}{r^2 \left( 1-\frac{2M}{ar}\right)[\frac{2M^2}{a} +\frac{2 M^2}{a^2}-\frac{2Mr}{a}  -\beta_oQ^2 \ln\left( 1- \frac{2M}{ar} \right)]},
\end{equation}  
which allows one to compute the radius of the innermost stable circular orbit
(ISCO).\\
Once the metric and the four-velocity of matter orbiting in the disk
are known it is possible to reconstruct the shape of the Fe line, as in
\citet{Psaltis_Johannsen12a}. Given an observer that sees the NS-disk system at an
inclination $\psi$ (the angle between the observer direction and the
perpendicular to the disk plane) light rays are traced from an image plane at the location of the
observer, until they reach the disk (or until
the intercept the surface of the NS in those cases, and for those
inclinations, for which the NS can occult/truncate the disk). Then one can
reconstruct the shape of the line, by integrating over the image plane (with
coordinates $[\eta,\zeta]$), the intensity due to the emission of the
disk. Ray-tracing maps each point $[\eta,\zeta]$ of the image plane to a
point on the equatorial plane where the disk is located. For each
point we can compute a transfer function that maps the frequency  of the
emitted photon $\nu_o$ to that of the observed photon $\nu$
according to $\nu/\nu_o = (k_\nu u_{\rm obs}^\nu)/(k_\nu u_{\rm
  disk}^\nu) = F$, where $k_\nu$ is the photon wave four-vector (either at
the position of 
the observer or of the emitter in the disk) whose value is provided by
the geodesic equations of the ray-tracing, while
$u_{\rm obs}^\nu$ and $u_{\rm disk}^\nu$ are respectively the four velocity of the observer (taken
at rest) and of the matter in the disk. The intensity $I_{\rm obs}$
at the observer position can be
computed, once the intensity of the radiation emitted in the disk
$I_{\rm disk} $ is
known, recalling that $I_{\rm obs}/(k_\nu u_{\rm
  obs}^\nu)^3 = I_{\rm disk}/(k_\nu u_{\rm
  disk}^\nu)^3$.
Then, the spectrum can be derived integrating the intensity over the
plane $[\eta,\zeta]$ at the observer location:
\begin{equation}
I(\nu) = \int  I_{\rm obs}(\eta,\zeta)\delta(\nu -\nu_o)F(\eta,\zeta)       d\eta d\zeta
\end{equation}  
In general one assumes that there is no emission coming from regions
inside the ISCO, while in the disk the emissivity scales as a
power-law of the
circumferential radius, $r_c^\gamma = r^\gamma
\tilde{g}_{\phi\phi}^{\gamma/2}$, where the equality comes from the
definition of the circumferential radius itself. A typical value is
$\gamma =-3$, and we use it in the following. The dependence on the radius is then steep enough that one
can truncate the disk emission around a few ISCO radii without
affecting the shape of the line.

\section{Results}
\label{sec:ray}

Given that we do not want to select a specific equation of state, or
scalar field coupling, to keep the discussion as general as possible
we treat the NS mass $M$, its radius $R$ and the total scalar charge
$Q$ as independent quantities. This is not true, given that those
three quantities are strongly related. This relation, however, is
 non trivial. Moreover, leaving these three quantities free, the result
can be easily applied to any STT-NS model. We chose
$\alpha_o=0$, and $\beta_o =-6$. Lowering $\beta_o$ to values around
the limit for spontaneous scalarization does not substantially modify
the results.

\begin{figure}
   	\centering
         \includegraphics[scale=0.5,clip,bb=10 33 450 300]{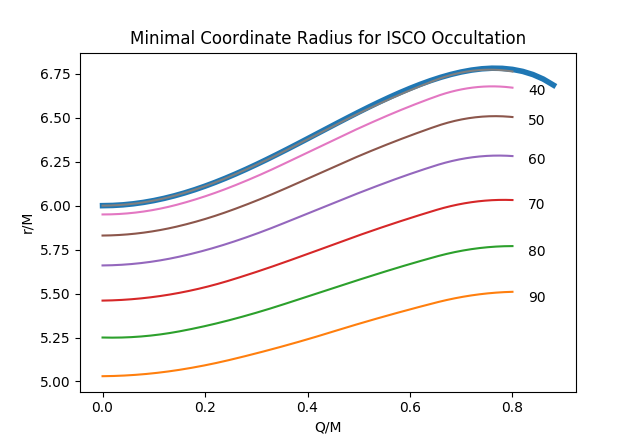}\\
         \includegraphics[scale=0.5]{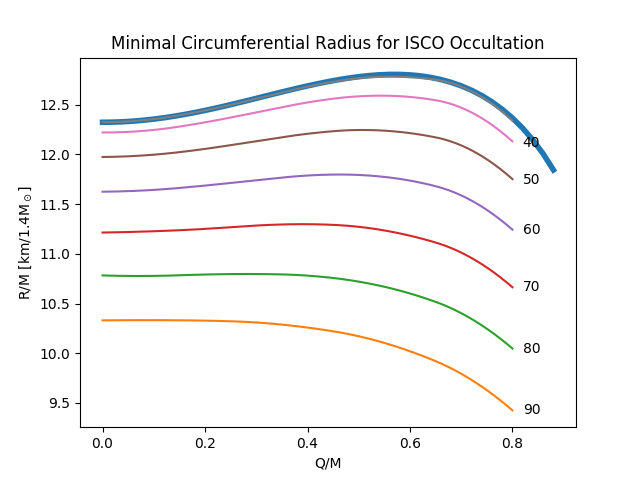}
 \caption{Minimal coordinate radius (upper panel) and circumferential
   radius (lower panel) for ISCO occultation by a NS as a function of
   the ratio of total scalar charge over the mass ($Q/M=0$ is GR), for various inclination of
 the observer with respect to the normal of the disk plane. The thick
 blue line is the ISCO radius as a function of the scalar charge.}
\label{fig:1}
 \end{figure}

Before investigating how STTs, and scalarized NSs, change the shape of the Fe line, we
begin by discussing under what conditions one can have a NS that
causes occultation of the ISCO. In Fig.~\ref{fig:1} we show the
minimal coordinate radius, and the minimal circumferential radius (the
only invariant quantity that can be physically measured),
such that the NS occults the ISCO, for various
inclination angles, together with the radius of the ISCO itself. It is evident that occultation/truncation can take place only
if the inclination angle of the observer is $\psi >
30^\circ$. Interestingly this threshold does not depend on the
presence of a scalar field. In GR, for systems seen edge on, when the
inclination angle of the observer is $\psi =
90^\circ$, occultation/truncation of the ISCO requires the NS coordinate radius to
be $> 5M$. In STTs this threshold increases by about 10\% for a
scalar charge $Q=0.8M$. This difference between GR and STTs holds also
for different viewing angles. Instead, in terms of the
circumferential radius, we see that the threshold radius for occultation of scalarized
NSs is smaller for large inclinations, and marginally larger
approaching a viewing angle of $\psi \sim 30^\circ$. Given that one of the
most relevant effect of a scalar field on the structure of NSs, is that
scalarized NSs have larger circumferential radii than their GR
counterparts of the same gravitations mass, 
occultation might be a more common phenomenon in scalarized systems
than in GR. In particular, given that there is a mass
threshold for spontaneous scalarization, one would expect 
occultation to be substantially more frequent above this mass.

\begin{figure}
   	\centering
         \includegraphics[scale=0.5,clip,bb=10 5 450 330]{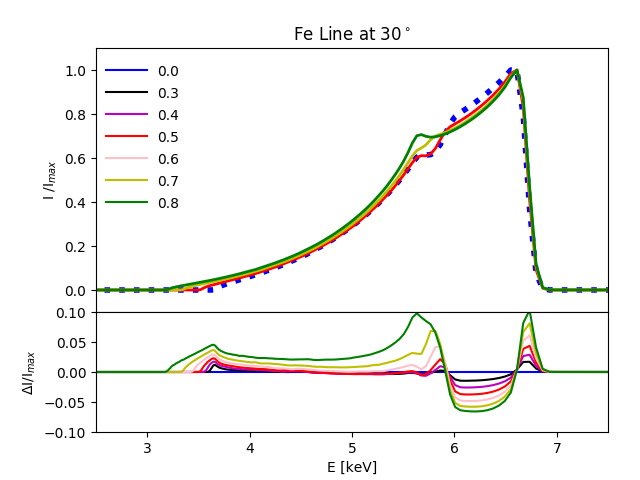}\\
 \caption{Upper panel: shape of the Fe K$_{\alpha}$ line, normalised
   to the maximum,  for a viewing angle of
   $30^\circ$, and different values of the ratio of total scalar charge over the mass. The thick dotted blue line is the GR case. Lower panel: percentage deviation of the line shape
   as a function of the scalar charge with respect to GR.}
 \label{fig:2}
 \end{figure}

 In Fig.~\ref{fig:2} we show the shape of the iron line for a viewing
 angle of $\psi =30^\circ$ in the absence of occultation, for various
 values of $Q/M$. It is intersting
 to note that the location of the edge at $\simeq 6.74$ keV does not
 depend on the presence of a scalar charge. On the other hand the
 effects of the scalar charge are more evident in the shape of the
 line. In particular, the intensity in the range $[5.9-6.6]$keV is
 smaller than in GR, from $\sim 2\%$ for $Q=0.3M$, to $\sim 7\%$ at
 $Q=0.8M$. For $Q>0.5M$ differences with respect to GR emerge also in
 the low energy tail. In particular we observe the formation of a
 \textit{second horn} at $\simeq 5.6$keV, and a tail which is about $2\%$
 brighter, and extends down to $3.2$keV, with respect to the low energy
 limit of $3.7$keV in GR. 
 
\begin{figure*}
   	\centering
         \includegraphics[scale=0.56,clip,bb=10 5 900 230]{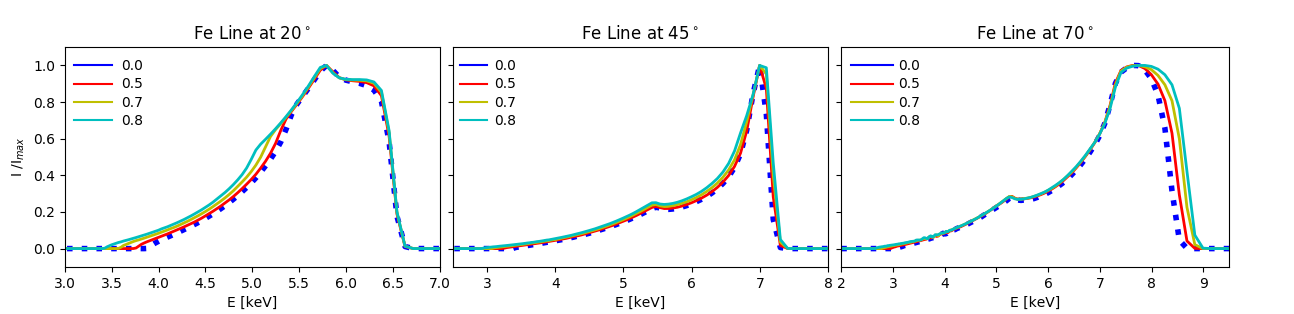}\\
 \caption{Shape of the Fe K$_{\alpha}$ line, normalised
   to the maximum,  for various viewing angles  
   ($20^\circ$ left,  $45^\circ$ center,
   $70^\circ$ right), and different values of the ratio of total scalar
   charge over the mass. The thick dotted blue line is the GR case.}
 \label{fig:3}
 \end{figure*}

 In Fig.~\ref{fig:3} we show the shape of the iron line for various
 viewing  angles, and for selected values of the scalar charge. It is
 evident that the way a scalar field modifies the line shape depends
 strongly on the viewing geometry. At $\psi =20^\circ$, the largest
 deviations are found in the intensity and shape of the low energy
 tail, and only partially in the shape of the $[6.0-6.5]$keV part. At
 $\psi =45^\circ$ instead the deviations are much smaller, while at
 $\psi =70^\circ$ they emerge again but now in the position of the high
 energy edge, which moves from $\simeq 8.2$keV to $\simeq 8.7$ keV,
 while the rest of the line shape is unaffected. The reason for this
 change with viewing angle is due to the fact that, for small viewing
 angles the shape of the line is mostly affected by gravitational
 redshift, and light bending, whose effects are more prominent in the
 low energy tails. This is where deviations from GR have the largest
 impact. On the other hand, when the inclination rises, and the disk is
 progressively seen more edge on, special relativistic effects due to
 orbital motion, and the related Doppler boosting, become
 dominant. The shape of the line now is more a tracer of the location
 and dynamics of the ISCO, which impacts mostly the high energy part of
 the line and the location of the edge. In general deviations in
 the intensity in the body of the line are small, at most few
 percent.

\begin{figure}
   	\centering
         \includegraphics[scale=0.50,clip,bb=10 5 900 340]{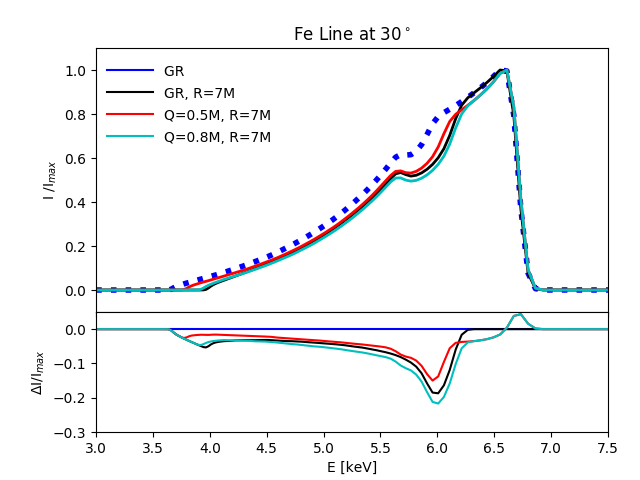}\\
 \caption{Shape of the Fe K$_{\alpha}$ line for  $30^\circ$ viewing
   angle, and various values of the scalar charge and NS radius
   normalised
   to the maximum (thick dotted blue line is unocculted GR). Lower panel percentage deviation of the line shape
   as a function of the ratio of total scalar charge over the mass, with respect to unocculted GR. }
 \label{fig:30occ}
 \end{figure}

In terms of occultation, for an observer inclination of $\psi =70^\circ$,
the effect are small (less than few percent) and mostly concentrated
in the low energy part of the line. However, when the NSs radius
become larger than the ISCO (or in case the disk is truncated at radii
larger than the ISCO radius), the high energy edge of the line begins to
move to lower energy. For $Q=0.8M$ and $R=7M$ (greater than the ISCO
radius for any $Q/M$) the leading edge is
located at $\simeq 8.4$keV. For an observer inclination of $\psi  =30^\circ$,
when occultation does not take place, the effect of a NS that
truncates the disk at radii
larger than the ISCO, is mostly concentrated in the intermediate part
of the line. Again the differences between the GR case, and STT are
only few percents, as shown in Fig.~\ref{fig:30occ}.

\section{Conclusions}
\label{sec:conc}

In this work we have investigated how the space-time deviations
produced by a non-minimally coupled scalar field, as hypothesised in some
alternative theory of gravity, could be probed using the shape of the Fe K$_{\alpha}$
line in accreting NS systems. Given that STTs satisfy the Weak
Equivalence Principle, standard  ray tracing techniques of GR can easily
be applied. The presence of a scalar field affects the shape of the
line in two ways: on one hand, it changes the space-time, affecting the
gravitational redshift and light bending;  on the other, it modifies
the Keplerian dynamics of matter orbiting in the disk, and the
location of the ISCO, which leads to further deviations in the line
shape associated to special relativistic Doppler boosting.

We found that such deviations however are at most a few percent, and
only for large total scalar charges $Q > 0.5M$. However, the
typical luminosity of Low Mass X-ray Binaries, where the Fe line has been detected, is
usually a sizeable fraction $0.05-0.11$ of the Eddington luminosity, and
 the intensity of the Fe line is typically 5-10\% of the
continuum. Given that the largest deviations from GR, in the line shape,
  extend over typical energy ranges $\sim 0.3-0.5$~keV, as can be seen, for example, from
  Fig.~\ref{fig:2}, and requiring the signal associated to these
  deviations to be well above ($S/N\sim 5$) the Poisson noise from the disk
  continuum (which, for isolated point-like sources, dominates the
  noise), in the same energy range, we can estimate the exposure time
  require to detect them. With the next generation of large collecting area X-ray
satellites like ATHENA (whose expected effective area at 6~keV is
$\simeq 2500$ cm$^2$, \citealt{Barcons_Barret+17a}) we predict that deviations in the intensity of
the line of the order of few percent could be detected with typical
exposure times ranging from $10^5$s in the brightest sources like Sco
X-1 e Ser X-1 to a few $10^5$s for weaker ones like 4U1608-52. More
interesting is the fact that for large viewing angles, the high-energy
edge can move enough to be revealed even with a low spectral
resolution. This, in our opinion, could be the easiest deviation
to measure. 

There are of course several other issues, that can play a role in the
correct modelling of the line shape \citep{Miller07a,Dauser_Garcia+16a}. It is well known that the choice
of the illuminator for example can affect it. The correct
modelling of the background plays also a crucial role, as well as the
presence of other lines that can blend \citep{Iaria_DAi+09a}. Not to talk about
the assumption of a disk truncated at the ISCO. It is also possible that scalarized NSs, having in
general larger radii than in GR  \citep{Damour_Esposito-Farese93a}, could lead to stronger occultation
effects. However, even if this could provide an
alternative way to measure the radius of the NS, it is not clear how
degenerate the information it provides is with respect to the equation
of state.

We stress again that this paper was organised as a proof of principle, and we opted for the
simplest possible approach to test the viability of this effect. Given
however the potential of this kind of measure as a possible independent test of
GR and its alternatives, we deem that a more accurate evaluations of
the expected results, considering specific STTs, or more
realistic EoS, to account for the mutual relation between mass, scalar
charge and NS radius, and specifically
targeted to known systems, is worth a further analisys.

\section*{Acknowledgements}
The authors wish to thank Riccardo La Placa, Luigi Stella, and Pavel
Bakala, for having pointed to us the possible use of iron lines in neutron stars as
tracer of the neutron star radii, and the feasibility of percentage measures on
the line shape with future X-ray satellites. The authors also
acknowledge financial support
from the ``Accordo Attuativo ASI-INAF n. 2017-14-H.0 Progetto: on the
escape of
cosmic rays and their impact on the background plasma'' and from the
INFN Teongrav collaboration. We finally thanks the referee D.~Ayzenberg for his
positive review.

\bibliography{MySTT}{}
\bibliographystyle{mn2e}

\end{document}